\documentclass{PoS}

\usepackage{amsmath}
\usepackage{microtype}

\title{EPPS16 -- First nuclear PDFs to include LHC data}

\ShortTitle{EPPS16 -- First nuclear PDFs to include LHC data}

\author{Kari J. Eskola\\
       University of Jyvaskyla, Department of Physics, P.O. Box 35, FI-40014 University of Jyvaskyla, Finland \\
       Helsinki Institute of Physics, P.O. Box 64, FI-00014 University of Helsinki, Finland \\
       E-mail: \email{kari.eskola@jyu.fi}}

\author{\speaker{Petja Paakkinen}
        \\ University of Jyvaskyla, Department of Physics, P.O. Box 35, FI-40014 University of Jyvaskyla, Finland \\
        E-mail: \email{petja.paakkinen@jyu.fi}}

\author{Hannu Paukkunen\\
       University of Jyvaskyla, Department of Physics, P.O. Box 35, FI-40014 University of Jyvaskyla, Finland \\
       Helsinki Institute of Physics, P.O. Box 64, FI-00014 University of Helsinki, Finland \\
       E-mail: \email{hannu.paukkunen@jyu.fi}}

\author{Carlos A. Salgado\\
       Instituto Galego de F\'{\i}sica de Altas Enerx\'{\i}as (IGFAE), Universidade de Santiago de Compostela, E-15782 Galicia, Spain \\
       E-mail: \email{carlos.salgado@usc.es}}

\abstract{
We present results of our recent EPPS16 global analysis of NLO nuclear parton distribution functions (nPDFs). For the first time, dijet and heavy gauge boson production data from LHC proton--lead collisions have been included in a global fit. Especially, the CMS dijets play an important role in constraining the nuclear effects in gluon distributions. With the inclusion of also neutrino--nucleus deeply-inelastic scattering and pion--nucleus Drell--Yan data and a proper treatment of isospin-corrected data, we were able to free the flavor dependence of the valence and sea quark nuclear modifications for the first time. This gives us less biased, yet larger, flavor by flavor uncertainty estimates. The EPPS16 analysis indicates no tension between the data sets used, which supports the validity of collinear factorization and universal nPDFs for nuclear hard-collision processes in the kinematical range studied.
}

\FullConference{The European Physical Society Conference on High Energy Physics\\
                 5-12 July\\
                 Venice, Italy}

\begin{document}

\section{Overview}

We discuss here the EPPS16 global analysis of next-to-leading order (NLO) nuclear parton distribution functions (nPDFs) \cite{Eskola:2016oht} with emphasis on the LHC proton--lead data, now included for the first time in an nPDF fit. We compare the results with those of other NLO nPDF analyses performed in past years.
In EPPS16, we have taken a step ahead from our earlier EPS09 analysis~\cite{Eskola:2009uj} by including a good number of completely new data types, namely the neutrino--nucleus deeply-inelastic scattering (DIS), pion--nucleus Drell--Yan (DY) and the LHC dijets and electroweak bosons, thus having more constraints than any other concurrent analysis.
Also, the number of data points has almost doubled from EPS09.
With these new constraints we have been able to
make the analysis more data-driven and less biased by allowing for a flavour freedom in the quark sector, and improve the determination of the nuclear gluon distributions.

\section{Analysis details}

\begin{floatingfigure}
  \vspace{-2ex}
  \includegraphics[width=0.51\textwidth]{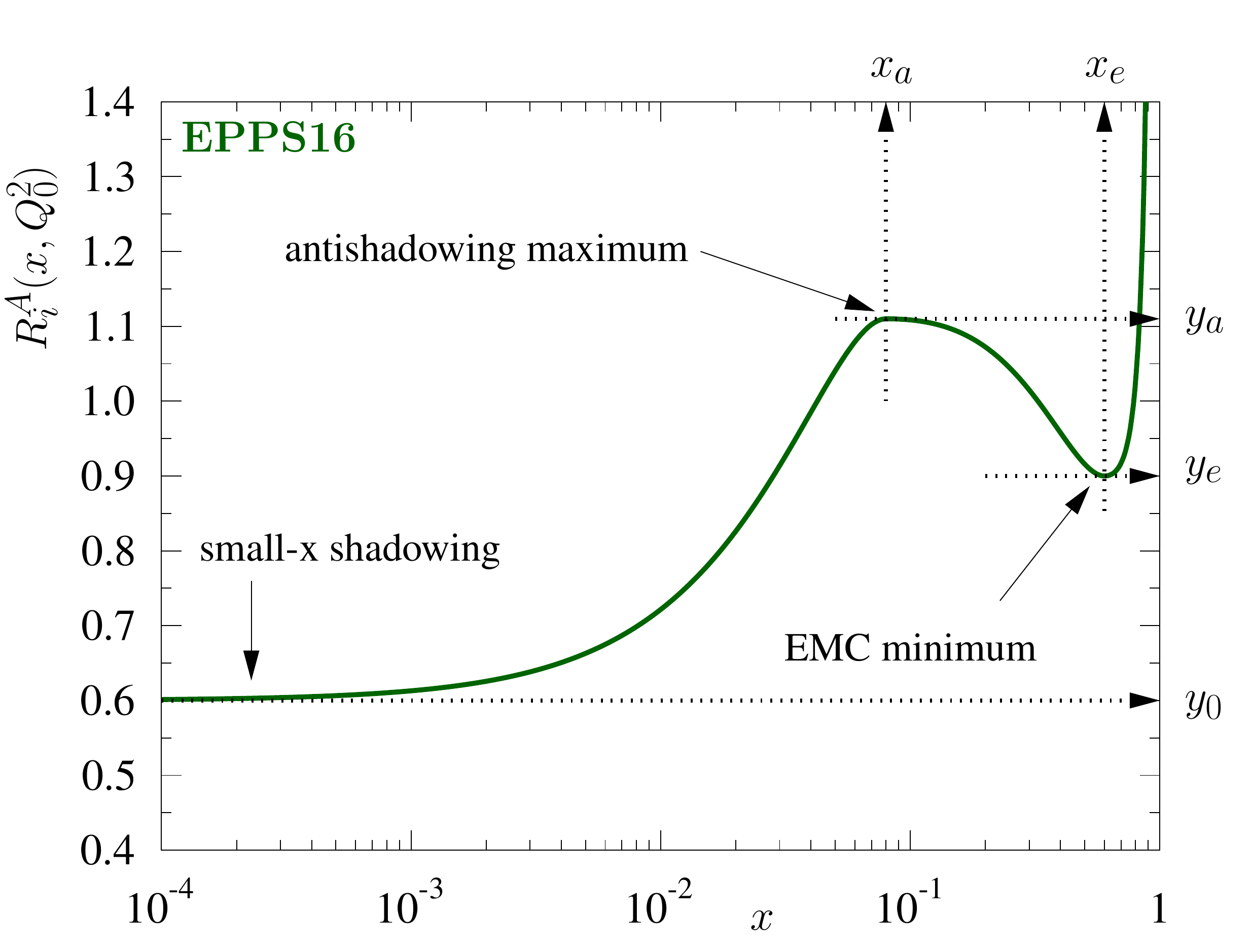}
  \caption{Functional form of the EPPS16 parametrization. Figure from Ref.~\cite{Eskola:2016oht}.}
  \label{fig:fit-form}
  \bigskip
\end{floatingfigure}

In our framework \cite{Eskola:2016oht} we parametrize the nuclear modification functions $R_i^A(x,Q^2)$ which are to be multiplied with the free proton PDFs $f_i^{{\rm p}}(x,Q^2)$, taken here to be those of CT14~\cite{Dulat:2015mca}, to obtain the bound nucleon distributions of parton flavour $i$,
\begin{equation}
  f_i^{{\rm p}/A}(x,Q^2) = R_i^A(x,Q^2) f_i^{{\rm p}}(x,Q^2).
\end{equation}
The parametrization is done at the charm-mass tresshold $Q_0^2 = m_c^2$ in terms of the location and height of the small-$x$ shadowing, antishadowing maximum and the EMC minimum as indicated in Figure~\ref{fig:fit-form}; the large-$x$ Fermi-motion slope is held fixed for all flavours.
The mass number dependence is parametrized with a power law behaviour in such a way that larger nuclei have larger modifications.

Most of earlier analyses have assumed identical modifications separately for valence and for sea quarks at the parameterization scale $Q^2_0$. We, for the first time, allow parametric freedom for all flavours,
\begin{equation}
  R_{\textcolor{black}{u_{\rm V}}}^A(x,Q^2_0) \neq R_{\textcolor{black}{d_{\rm V}}}^A(x,Q^2_0),
  \qquad
  R_{\textcolor{black}{\bar{u}}}^A(x,Q^2_0) \neq R_{\textcolor{black}{\bar{d}}}^A(x,Q^2_0) \neq R_{\textcolor{black}{\bar{s}}}^A(x,Q^2_0).
  \label{eq:flavoursym}
\end{equation}
The latest nCTEQ analysis~\cite{Kovarik:2015cma} has flavour freedom for valence, but not for sea quarks.
Now that we allow the initial flavour freedom in the fit, it is important to undo the isospin corrections used in the published DIS data. These were introduced by the experiments to ease the interpretation of the data, but from the global analysis viewpoint such corrections are merely complications and sources of bias.
Also, since the neutrino DIS and the LHC measurements lack a baseline proton or deuterium measurement, we include these either normalized or as forward-to-backward ratios to reduce experimental systematic uncertainties and sensitivity to the free-proton PDFs.
Where applicable, we take into account the correlated systematics.
Importantly, we have also removed all the data weights which were still present in the EPS09 analysis.
For the uncertainty analysis we use the standard Hessian method with a global tolerance $\Delta\chi^2 = 52$ such that on average for any chosen error set, all data sets remain within their 90\% confidence ranges, see Ref.~\cite{Eskola:2016oht} for details.

\begin{floatingfigure}
  \vspace{-2ex}
  \includegraphics[width=0.51\textwidth]{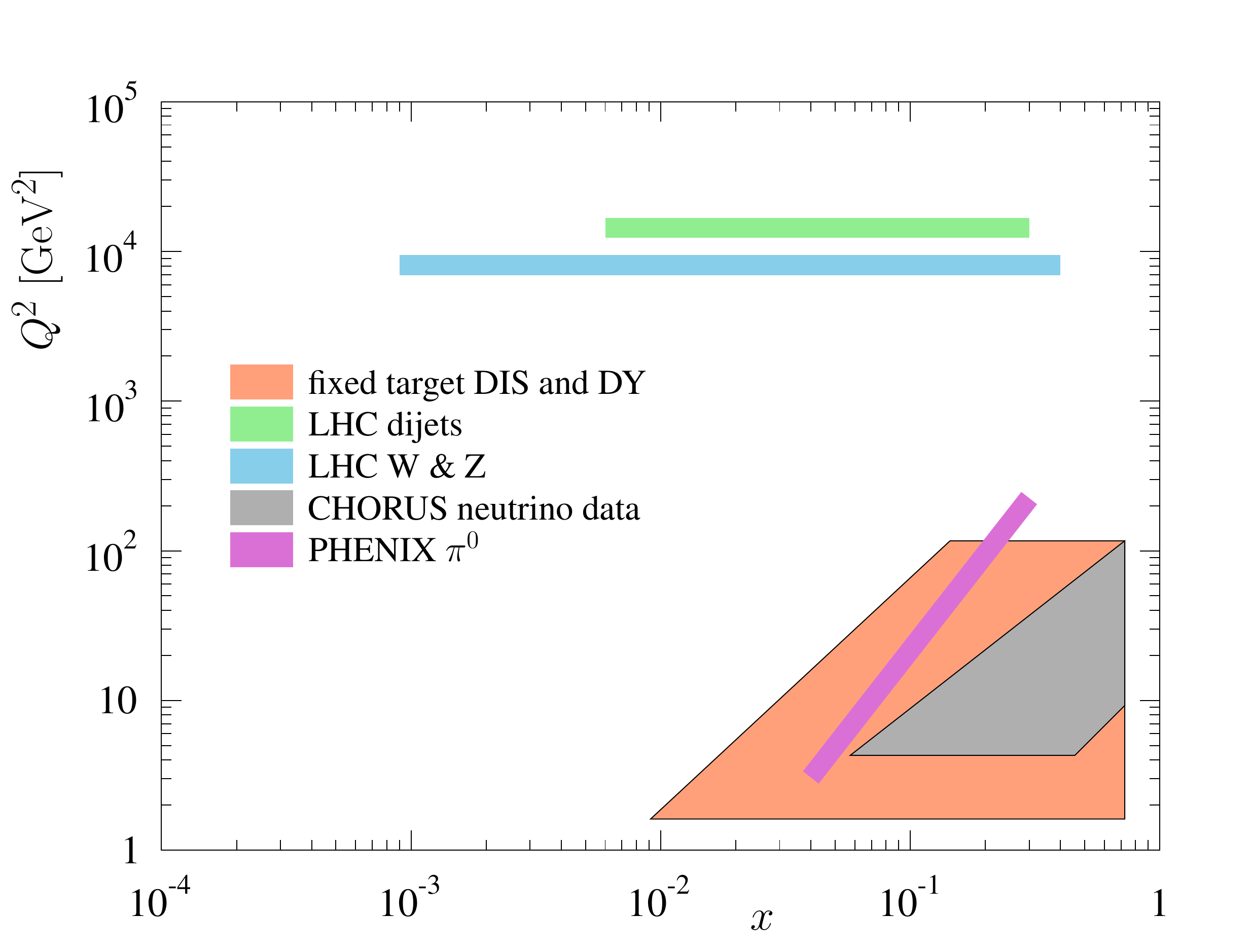}
  \caption{Kinematical reach of the data used in EPPS16. Figure from Ref.~\cite{Eskola:2016oht}.}
  \label{fig:xQ2}
  \bigskip
\end{floatingfigure}

\section{New experimental input}

\begin{floatingfigure}[b]
  \hspace{-0.18cm}
  \begin{minipage}{0.333\textwidth}
    \includegraphics[width=\textwidth]{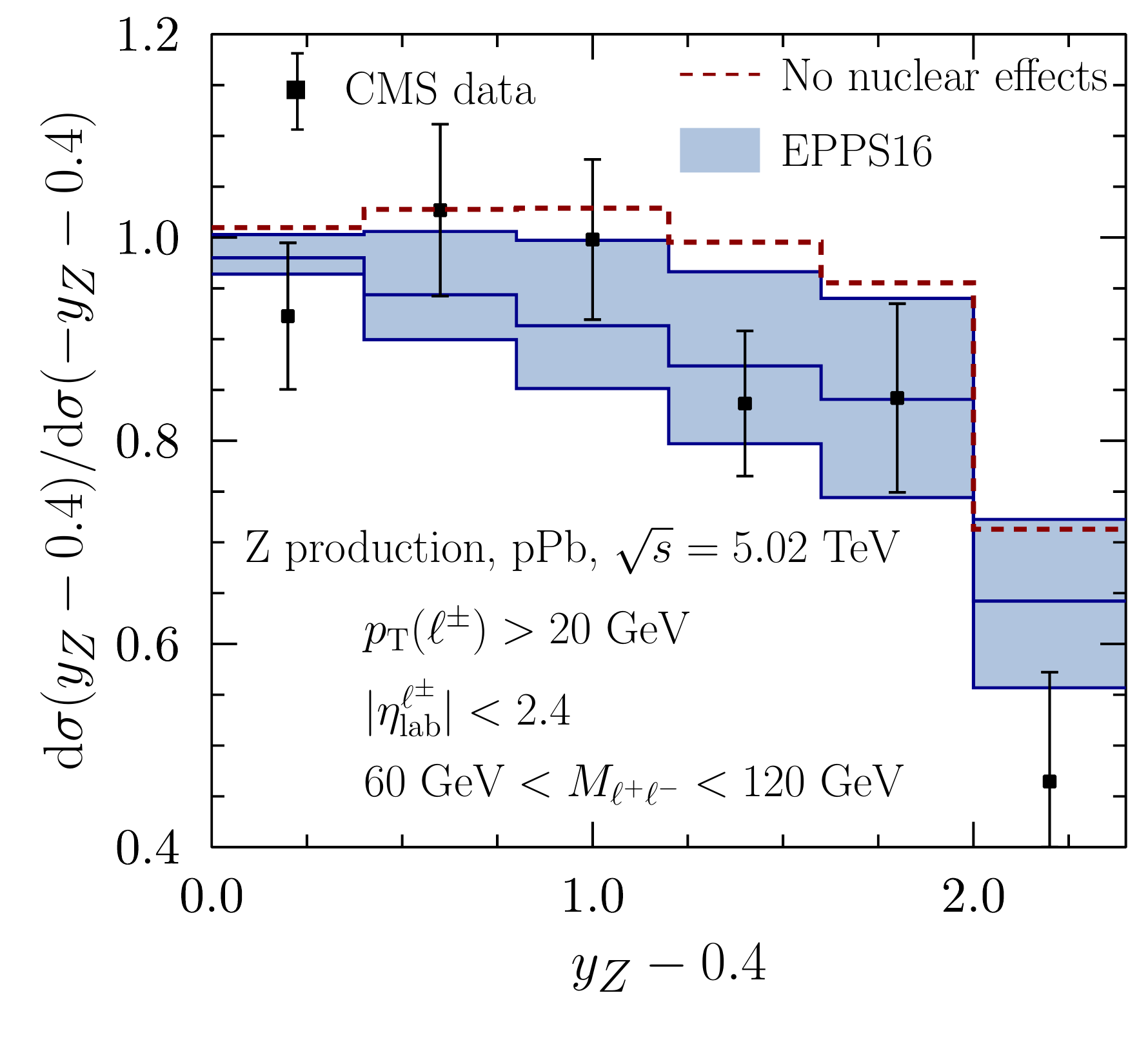}
    \includegraphics[width=\textwidth]{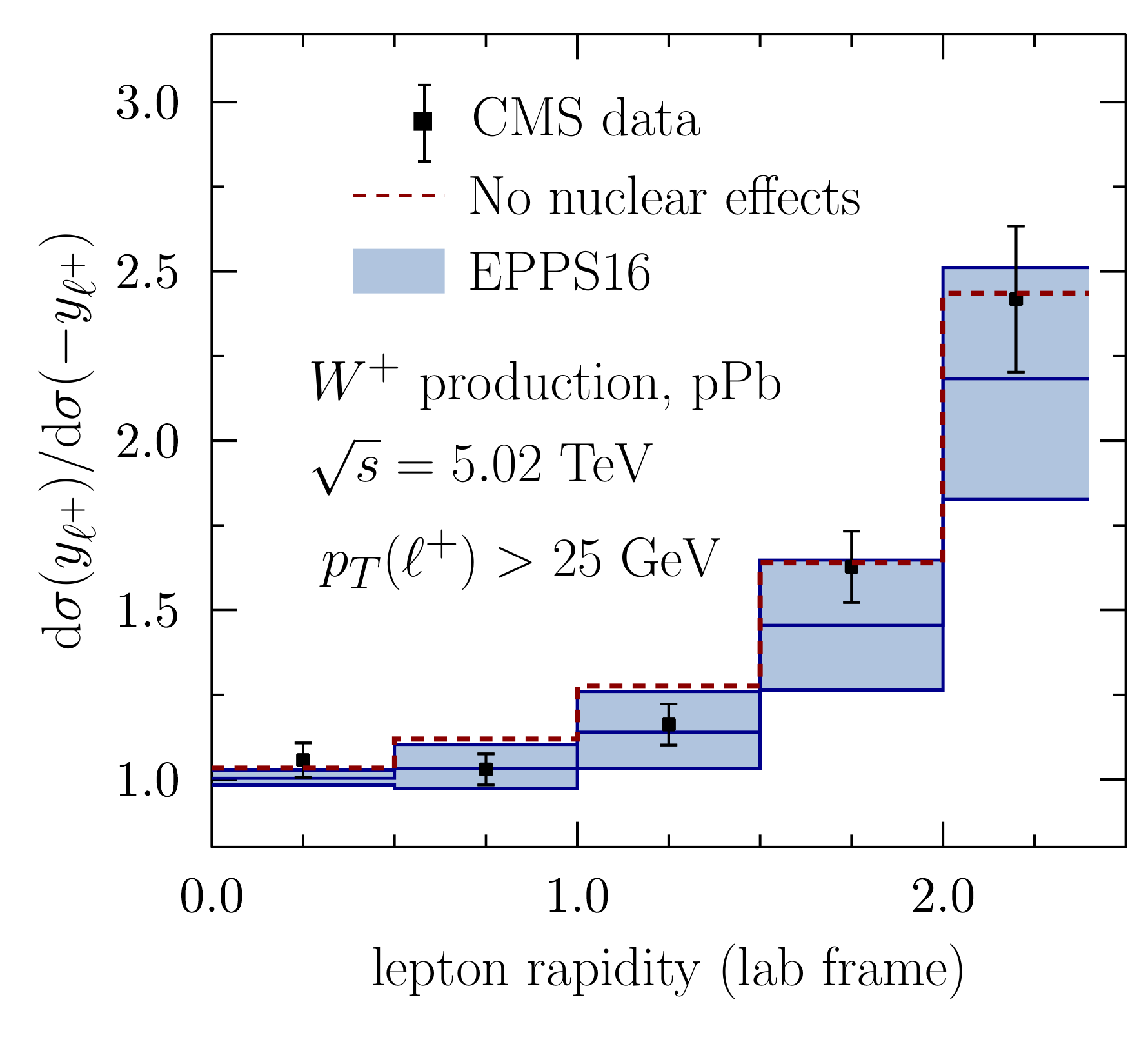}
  \end{minipage}
  \hspace{-0.18cm}
  \begin{minipage}{0.333\textwidth}
    \includegraphics[width=\textwidth]{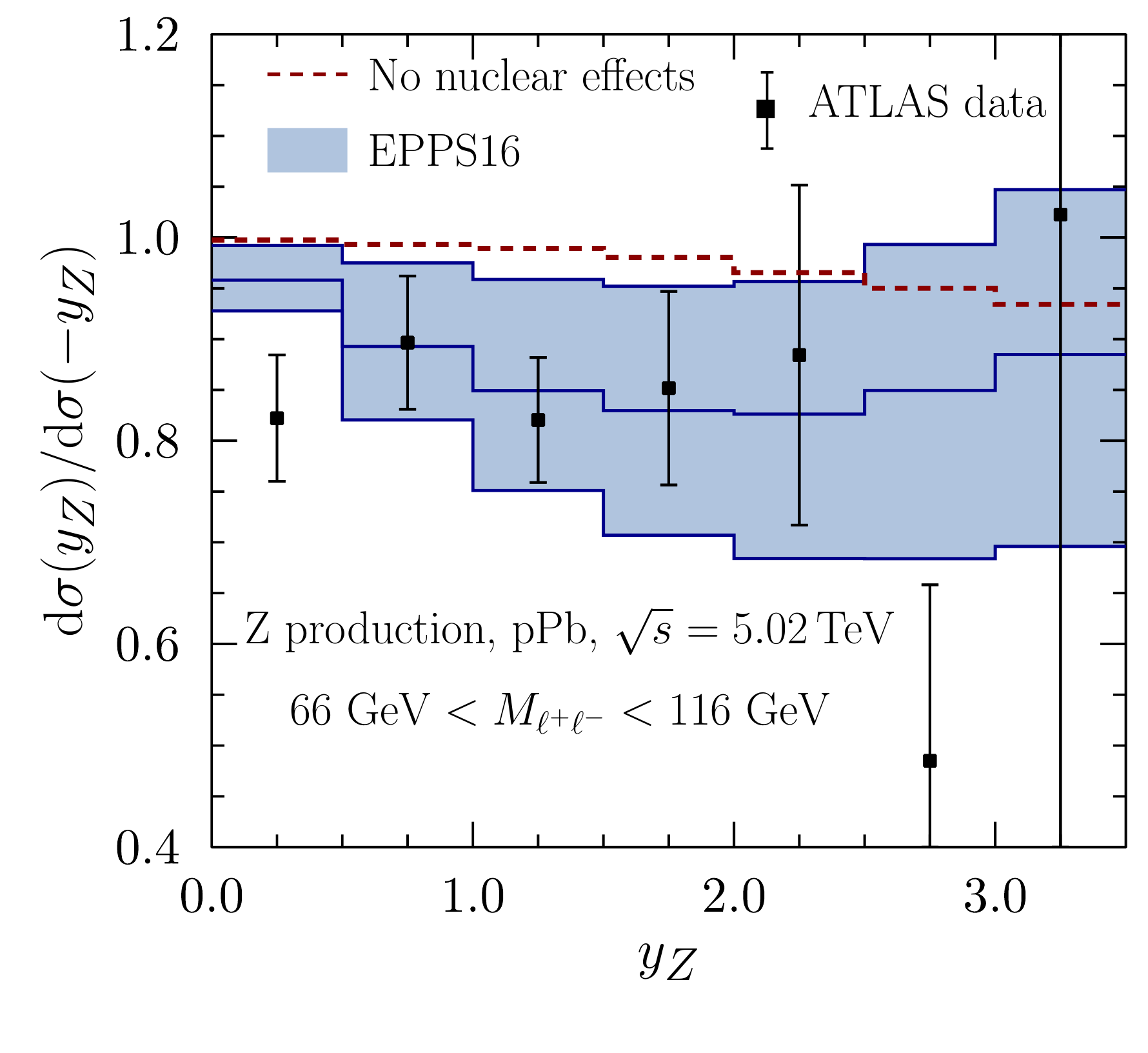}
    \includegraphics[width=\textwidth]{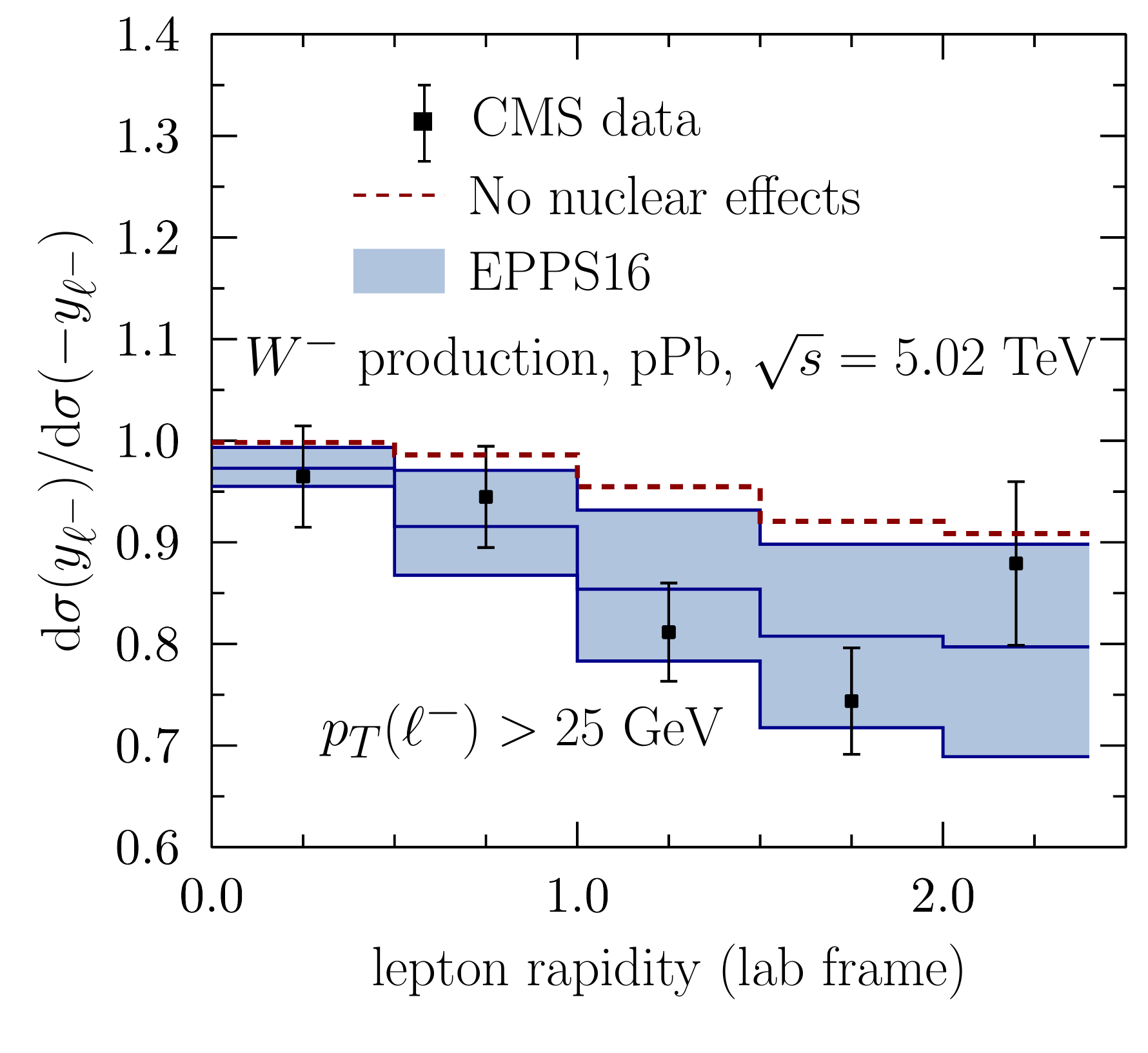}
  \end{minipage}
  \hspace{-0.18cm}
  \begin{minipage}{0.333\textwidth}
    \includegraphics[width=\textwidth]{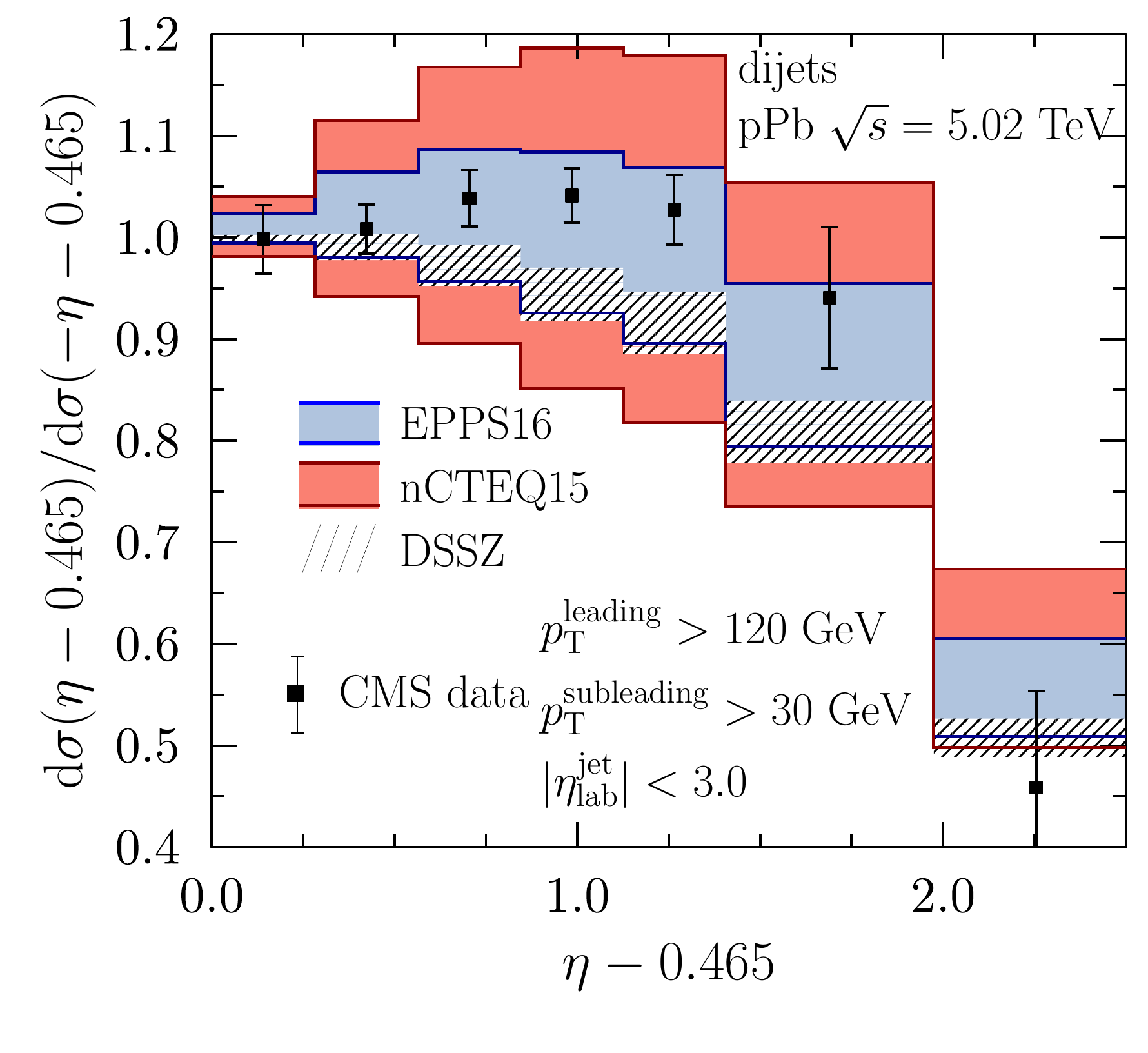}
  \end{minipage}
  \hspace{-0.18cm}
  \vspace{-1ex}
  \caption{The LHC 5.02 TeV pPb data used in the EPPS16 analysis: Z production~\cite{Khachatryan:2015pzs,Aad:2015gta} (upper left and middle panels), W production~\cite{Khachatryan:2015hha} (lower left and middle panels) and dijets~\cite{Chatrchyan:2014hqa} (rightmost panel). Figures from Ref.~\cite{Eskola:2016oht}.}
  \label{fig:LHC}
\end{floatingfigure}

The LHC data open a whole new, high-$Q^2$, kinematic region for nPDF studies, see Figure~\ref{fig:xQ2}. The LHC data used in the EPPS16 analysis are shown in Figure~\ref{fig:LHC} with the corresponding EPPS16 fit results.
For the Z production~\cite{Khachatryan:2015pzs,Aad:2015gta} we obtain a good fit, with the data supporting net nuclear shadowing at small $x$, but the obtainable constraints are limited by low statistics. For W~\cite{Khachatryan:2015hha} the statistics are a bit better, but still more data are needed for better constraints.
With the scarce statistics and also due to losing some information in using the forward-to-backward ratios, the electroweak observables cannot be utilized to their full potential at the moment.
The new proton--proton baseline measurement at the same energy may improve the constraining power of these data in the future.

Of the LHC data, the most important part is here played by CMS dijets~\cite{Chatrchyan:2014hqa}. Without these data, the fit converges to parameter values indicating no gluon EMC effect, but once the dijet data are included, a clear preference for a gluon EMC slope is seen.
Similarly, the CHORUS neutrino--nucleus data~\cite{Onengut:2005kv} are important in constraining the flavour decomposition of nuclear modifications. The fit without neutrino data ends up to a situation where the u and d valence quark modifications differ significantly from each other, but when these data are included the fit converges to a parameter region with similar u and d valence modifications.
This is in accordance with the pion--nucleus DY data~\cite{Badier:1981ci,Bordalo:1987cs,Heinrich:1989cp}, which also seem to favour relatively similar modifications for both valence quarks~\cite{Paakkinen:2016wxk}. These data are now also for the first time included in a nPDF global analysis, but appear to have less constraining power than the neutrino--nucleus DIS.

\section{Results}

\begin{floatingfigure}[b]
  \includegraphics[width=\textwidth]{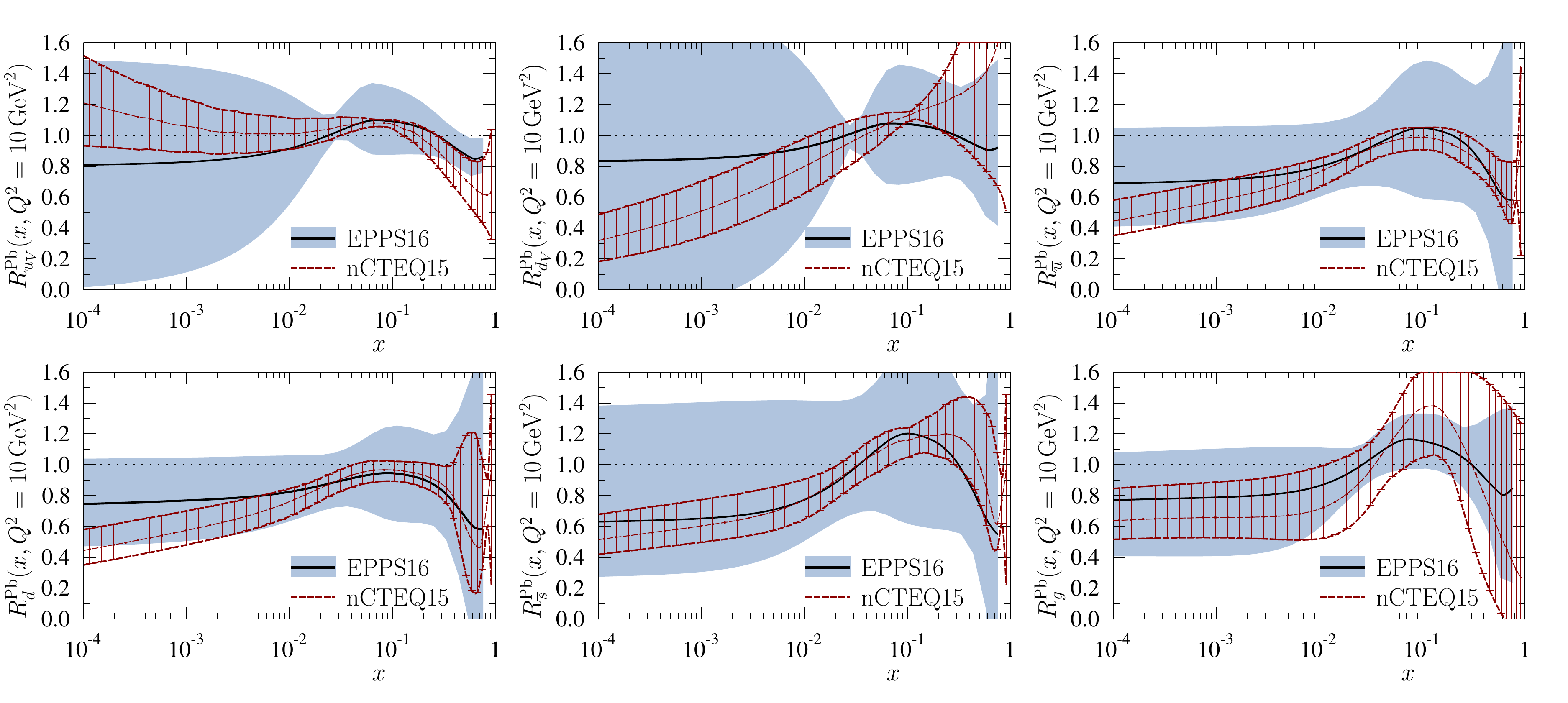}
  \vspace{-4ex}
  \caption{The EPPS16 and nCTEQ15 nuclear PDFs. Figure from Ref.~\cite{Eskola:2016oht}.}
  \label{fig:nCTEQ}
\end{floatingfigure}

The resulting nuclear modifications for lead nucleus at the scale $Q^2 = 10~{\rm GeV}^2$ are shown along with the respective nCTEQ15~\cite{Kovarik:2015cma} results in Figure~\ref{fig:nCTEQ}. The EPPS16 central best fit, shown as a black line, suggests a similar shape of modifications for all the flavours. That is, all the flavours experience small-$x$ shadowing, mid-$x$ antishadowing and high-$x$ EMC-effect.
This is contrary to the nCTEQ15 results, where the valence quarks exhibit quite different modifications. This is possibly due to nCTEQ15 using isospin-symmetrized DIS data and having no neutrino--nucleus DIS in the fit.
However, as the error bands always overlap, the EPPS16 and nCTEQ15 are compatible.

As seen in Figure~\ref{fig:nCTEQ}, we find the nCTEQ15 uncertainties to appear generally smaller than in EPPS16, which
is due to nCTEQ15 having less freedom in the fit.
For example, nCTEQ15 has only 2 free parameters for all the sea quarks, whereas EPPS16 has altogether 9.
The situation is quite different for mid-to-high-$x$ gluons, where nCTEQ15 has plenty of parameters but with harder $Q^2$ cut, and not including the dijet data, they have larger uncertainties.
This larger uncertainty translates as a wider uncertainty band in the case of dijets, see Figure~\ref{fig:LHC}, where the data (and EPPS16) clearly have smaller uncertainties.
All in all, the EPPS16 error bands are typically larger but less biased.


\begin{floatingfigure}[t]
  \hspace{-0.18cm}
  \includegraphics[width=0.333\textwidth]{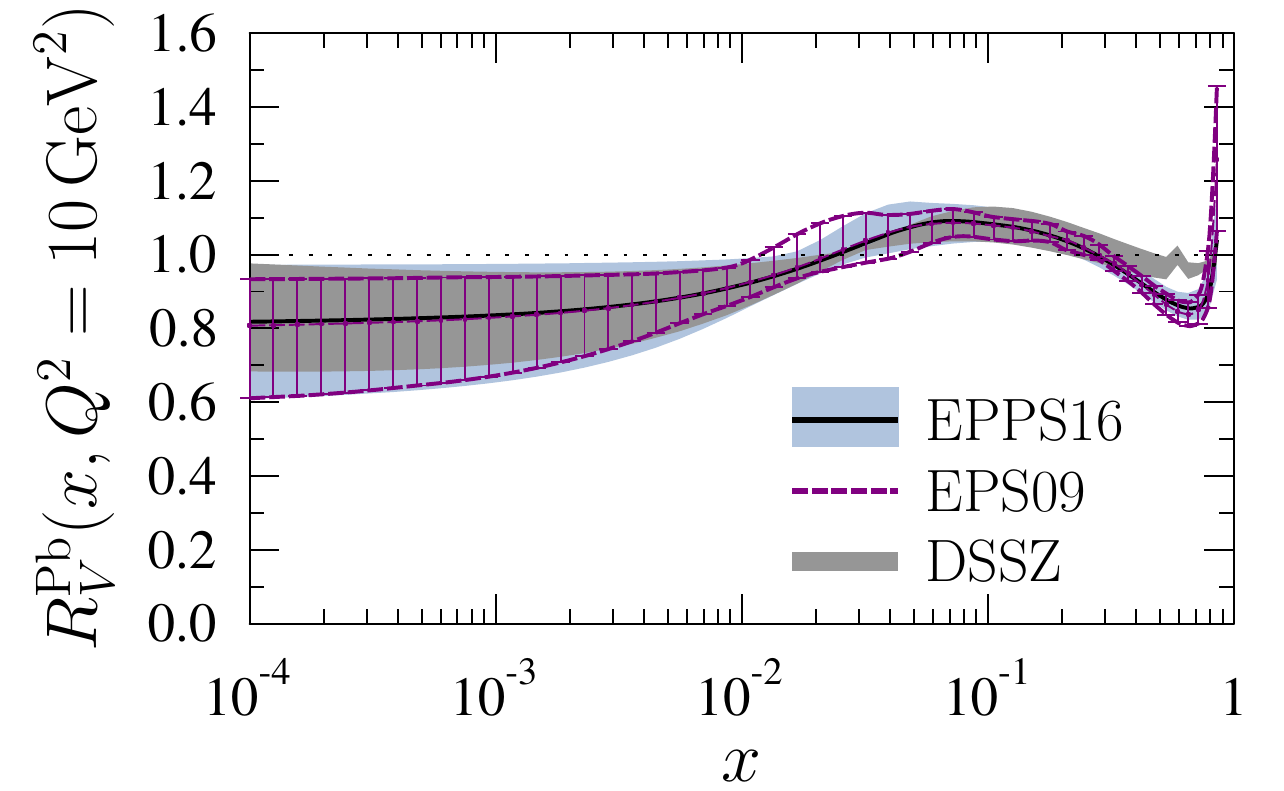}
  \hspace{-0.18cm}
  \includegraphics[width=0.333\textwidth]{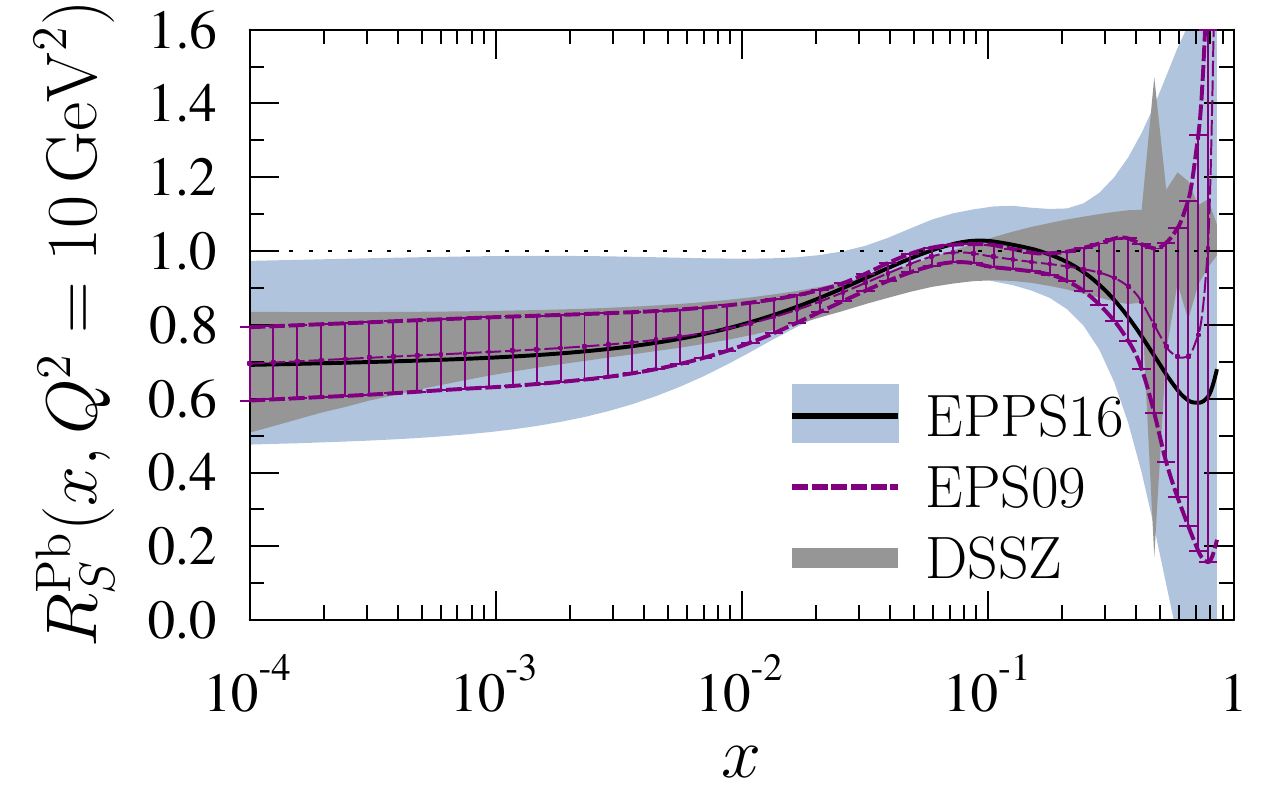}
  \hspace{-0.18cm}
  \includegraphics[width=0.333\textwidth]{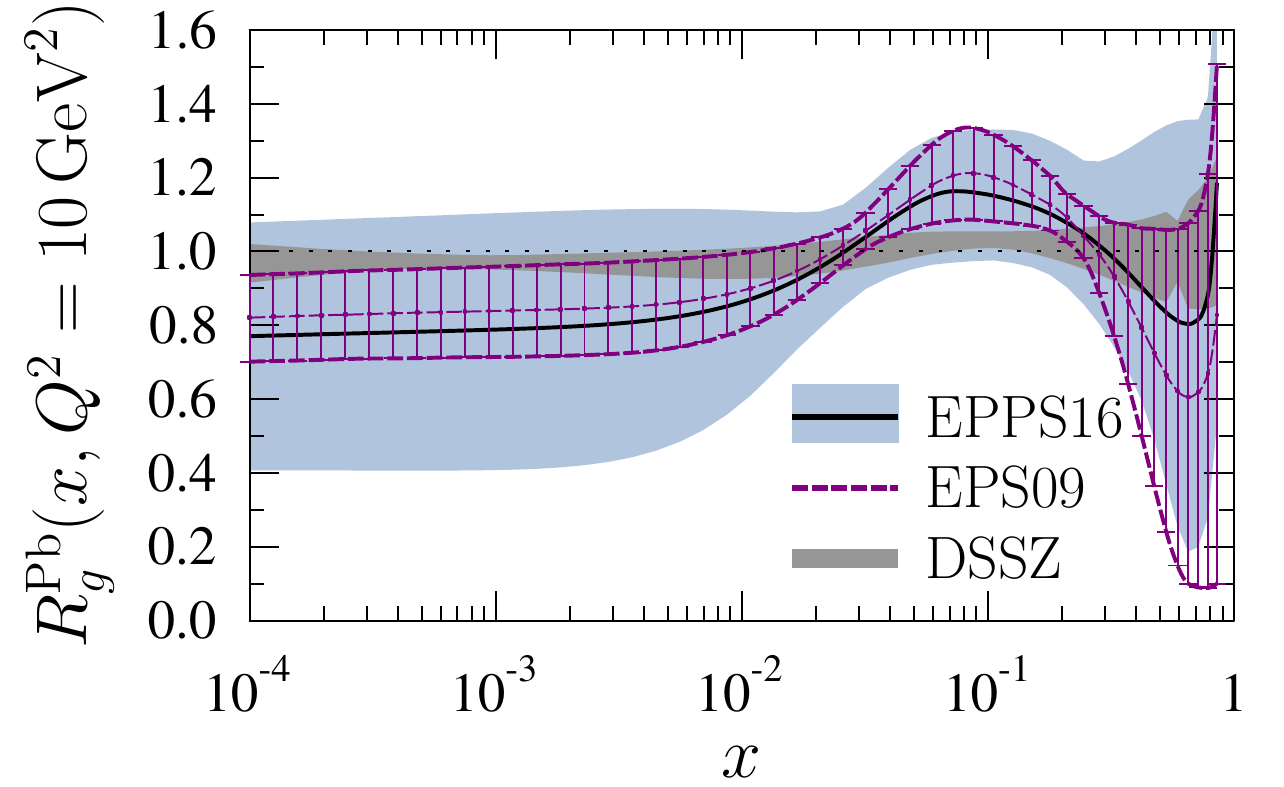}
  \hspace{-0.18cm}
  \vspace{-3ex}
  \caption{The EPPS16, EPS09 and DSSZ nuclear PDFs. Figure from Ref.~\cite{Eskola:2016oht}.}
  \label{fig:DSSZ}
\end{floatingfigure}

In Figure~\ref{fig:DSSZ} we compare the EPPS16 modifications to those of EPS09~\cite{Eskola:2009uj} and DSSZ~\cite{deFlorian:2011fp}. Since the latter two assume flavour symmetric modifications, we only compare the average modifications of valence and light sea quarks,
\begin{equation}
    R_{\rm V}^{\rm Pb} \equiv  \frac{u^{\rm p/Pb}_{\rm V}+d^{\rm p/Pb}_{\rm V}}{u^{\rm p}_{\rm V}+d^{\rm p}_{\rm V}},
    \qquad
    R_{\rm S}^{\rm Pb} \equiv  \frac{\overline{u}^{\rm p/Pb}+\overline{d}^{\rm p/Pb}+\overline{s}^{\rm p/Pb}}{\overline{u}^{\rm p}+\overline{d}^{\rm p}+\overline{s}^{\rm p}}.
\end{equation}
All three appear to be compatible with each other, except the large-$x$ DSSZ valence quarks, which likely suffer from ignoring the isospin corrections.
The EPPS16 uncertainties are larger for sea quarks and gluons due to having more freedom in the fit.
One should also note that the EPS09 gluon uncertainties are artificially small since an additional weight for the PHENIX data on inclusive pion production~\cite{Adler:2006wg} was used. DSSZ have practically no gluon modifications since they include nuclear modifications to fragmentation functions. Combined with CT14 proton PDFs, the nPDFs obtained with this choice seem to be in disagreement with the dijet data, as seen in Figure~\ref{fig:LHC}.

\section{Summary}

The LHC proton--lead data have opened a new high-$Q^2$ window for global nuclear PDF analyses. We have reported here the impact of these data on the EPPS16 nPDFs. The most important part is played by CMS dijets, which are essential in constraining the nuclear effects in gluon distributions. For electroweak observables the statistics are still insufficient to yield stringent constraints.
With the inclusion of neutrino--nucleus DIS and pion--nucleus DY data and a proper treatment of isospin-corrected data, we have been able to freely parametrize the flavor dependence of the valence and sea quark nuclear modifications. As a result, the EPPS16 uncertainties are generally larger but less biased compared to previous analyses.
We find that a consistent fit for a wide variety of observables in the kinematic range up to the electroweak scale can be achieved, which supports collinear factorization and universality of nPDFs.

\acknowledgments

We have received funding from the Academy of Finland, Project 297058 of K.J.E.\ and 308301 of H.P.; the European Research Council grant HotLHC ERC-2011-StG-279579; Ministerio de Ciencia e Innovaci\'{o}n of Spain and FEDER, project FPA2014-58293-C2-1-P; Xunta de Galicia (Conselleria de Educacion) - H.P.\ and C.A.S.\ are part of the Strategic Unit AGRUP2015/11. P.P.\ acknowledges the financial support from the Magnus Ehrnrooth Foundation.


\end{document}